\documentclass[prl,aps,groupedaddress,twocolumn,superscriptaddress,floatfix,showpacs]{revtex4}
\usepackage{epsfig}
\usepackage{mathrsfs}
\usepackage{amsmath}
\usepackage{textcmds}
\usepackage{amssymb}
\usepackage{mathptmx}
\allowdisplaybreaks[1]
\newcommand{\be}{\begin{equation}}
\newcommand{\ee}{\end{equation}}
\newcommand{\bea}{\begin{eqnarray}}
\newcommand{\eea}{\end{eqnarray}}
\newcommand{\ket}[1]{\left|#1\right\rangle}
\newcommand{\bra}[1]{\left\langle #1\right|}

\newcommand{\bc}{\begin{center}}
\newcommand{\ec}{\end{center}}

\renewcommand{\(}{\left(}
\renewcommand{\)}{\right)}
\renewcommand{\[}{\left[}
\renewcommand{\]}{\right]}
\newcommand{\forget}[1]{}

\newcommand{\re}{{\rm e}}

\newcommand{\ri}{{\rm i\,}}
\begin{document}
\title{Sub-nanoscale Resolution for Atom Localization, Lithography and Microscopy via Coherent Population Trapping}
\author{K. T. Kapale}
\email{KT-Kapale@wiu.edu}
\affiliation{Department of Physics, Western Illinois University, Macomb, Illinois, 61455-1367, USA}
\affiliation{Hearne Institute for Theoretical Physics, Department of Physics \& Astronomy,
Louisiana State University,
Baton Rouge, Louisiana 70803-4001, USA}
\author{G. S. Agarwal}
\affiliation{Department of Physics, Oklahoma State University, Stillwater, OK 74078, USA}
\begin{abstract}
We present a coherent population trapping based scheme to attain sub-nanoscale resolution for atom localization, microscopy and lithography. Our method uses three-level atoms coupled to amplitude modulated probe field and spatially dependent drive field. The modulation of the probe field allows us to tap into the steep dispersion normally associated with electromagnetically induced transparency and offers an avenue to attain sub-nanometer resolution using just optical fields.  We illustrate application of the techniques to the area of microscopy and lithography and show how multilevel schemes offer the possibility of improving resolution further.
\end{abstract}
\maketitle

{\it Introduction.}---The Rayleigh limit of resolution ($\lambda/2$), when using a light field of wavelength $\lambda$ for imaging or for lithographic fabrication, primarily arises due to the choice of the optical design that is limited by diffraction. It has been shown that the ultimate limit of resolution is the Heisenberg limit and within its bounds, in principle, one can obtain sub-Rayleigh resolution~\cite{Vigoureux:1992}. In general, sub-Rayleigh resolution can be achieved by employing diffraction-free techniques. Several different avenues are being explored on widely different fronts for obtaining sub-Rayleigh resolution based on near-field optics~\cite{Girard:1996}, quantum entanglement~\cite{Boto:2000}, monitoring fluorescence from atoms or molecules~\cite{Hell:2007}, and atomic coherence effects~\cite{PaspalakisKnight,AgarwalKapale:2006,HarrisFleischhauer,Arimondo:1996}. Nanoscale spatial resolution is useful for applications to the fields of lithography~\cite{KiffnerPark}, imaging~\cite{Li:2008}, atom localization~\cite{SahraiKapale}, high precision interferometry~\cite{Chatellus:2009} and it offers precise spatial selectivity for atomic qubits necessary for quantum information processing~\cite{Gorshkov:2008}. 

We have shown recently that coherent population trapping (CPT)~\cite{Arimondo:1996} can be used to localize atoms to subwavelength regimes~\cite{AgarwalKapale:2006}. The idea has been taken further by others to theoretically propose schemes for subwavelength microscopy~\cite{Yavuz:2007}, subwavelength patterning of Bose-Einstein condensates~\cite{Mompart:2009}, nanoscale trapping potentials for atoms~\cite{Yavuz:2009}, and two dimensional localization by coupling the atom with two spatially dependent fields~\cite{Jin:2009}. While the CPT based super-resolution techniques continue to be developed;  we note that a popular technique that attains nanometer resolution is the confocal fluorescence microscopy that makes use of the saturation of a two level transition. In another well-known, Stimulated Emission Depletion (STED), method one cuts down emission from outer regions by using strong fields~\cite{Hell:2007}. Fujita et al.~\cite{Fujita:2007} use the nonlinear relation between the excitation and fluorescence to demonstrate improved spatial resolution in three dimensions beyond the diffraction limit.

Thus the nanoscale resolution is becoming rather common in microscopic and other applications. The question is how to break this barrier. In this letter we show how CPT can be adopted to reach sub-nanoscale resolution. We use the physics behind CPT and Electromagnetically-Induced Transparency (EIT)~\cite{HarrisFleischhauer}. The sharp dispersion offered by EIT allows unprecedented control over the group velocity of the probe field. To take advantage of this steep dispersion we propose the use of amplitude modulated probe field and show that, in the best case scenario, the resolution can be further improved by a factor of 20 beyond that offered by the unmodulated CPT. The modulated CPT  gives resolution of 0.25 nm for the drive field of optical wavelength 500 nm. Thus we offer an avenue to attain sub-nanometer scale resolution which would have applications to atom localization, imaging and lithography.

The paper is organized as follows. First we briefly summarize the CPT-based atom localization scheme that is studied in detail in Ref.~\cite{AgarwalKapale:2006}. Then we introduce amplitude modulation for the weak probe field either as a perturbation or as a full modulation and study its effects on the localization. We present our results in terms of the point spread function of our scheme, which signifies the spread in the size of a point object as perceived by the scheme.

\begin{figure}[ht]
\centerline{\includegraphics[width=0.8\columnwidth]{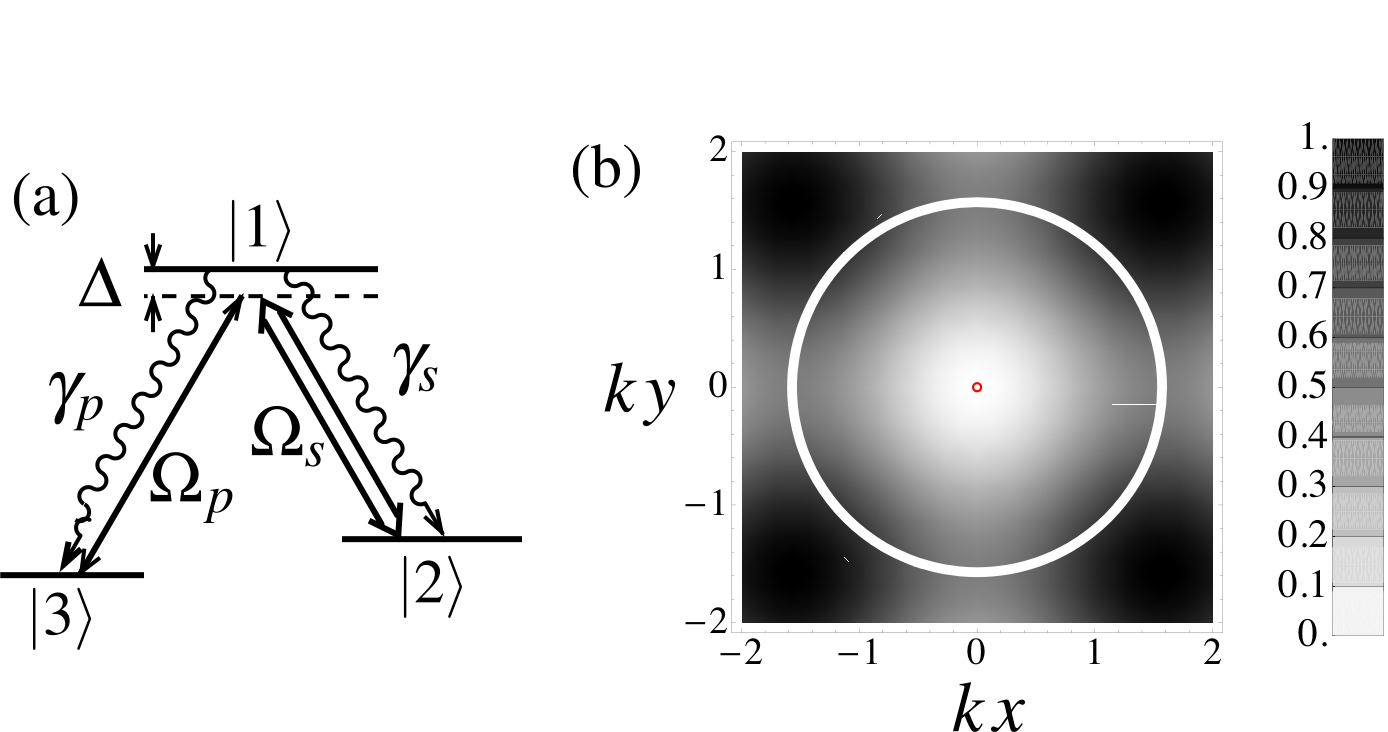}}
\caption{\label{Fig:Levelscheme} (Color online) (a) The atomic levelscheme for coherent population trapping. (b) Comparison of the measurable feature size within the Rayleigh limit (white circle) and with the CPT scheme (small red circle in the center) for $\mathcal{R} = 1000$ in the $x$-$y$ plane. At this value of $\mathcal{R}$ the improvement in resolution is by a factor of about 50.}
\end{figure}

\begin{figure}[ht]
\centerline{\includegraphics[width=\columnwidth]{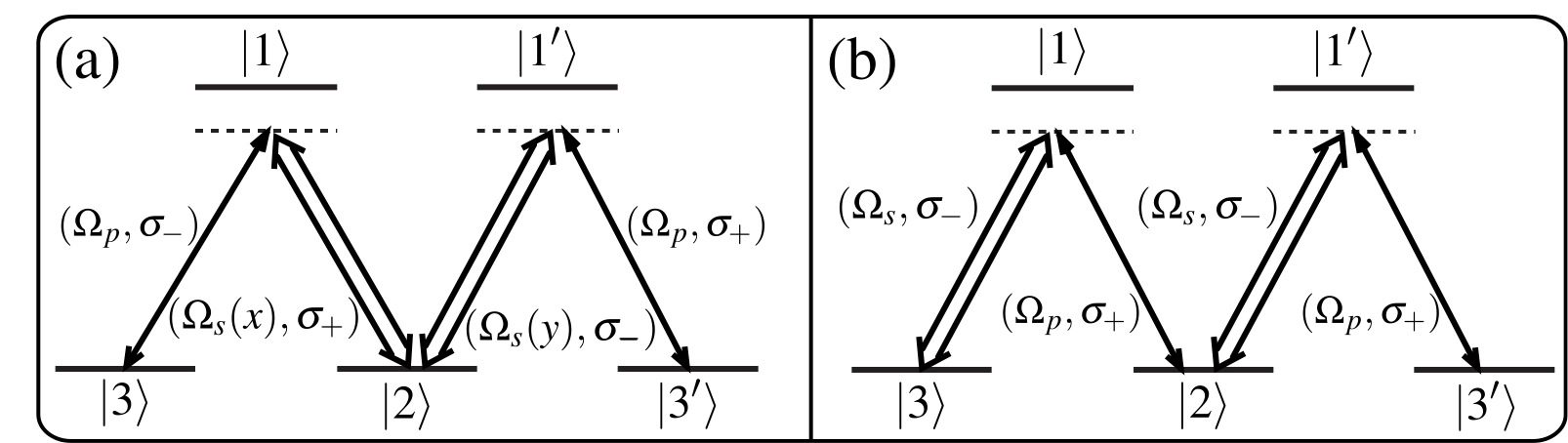}}
\caption{\label{Fig:MultiLevel} (a) Multilevel scheme for 2D imaging. The population of level 3 is $\rho_{33}(x,y)$ and shows sharp features. Notice the $x$ and $y$ dependance of the drive field coupling to level $\ket{3}$ (b) Multilevel scheme for further enhancement of resolution in a single dimension. This can be envisioned to be two coupled $\Lambda$ systems giving square of the Fabry-Perot function for $\rho_{11}(x)$.}
\end{figure}

{\it Model.}---We consider a three-level atom, as depicted in Fig.~\ref{Fig:Levelscheme}(a), illuminated by two optical fields, strong drive field on the $\ket{1}$-$\ket{2}$ transition  and weak probe field on $\ket{1}$-$\ket{3}$ transition with Rabi frequencies $\Omega_{s}$ and $\Omega_{p}$ respectively. We allow the fields to be detuned from their respective transitions but always assume two-photon resonance such that the two detunings are equal to a chosen value $\Delta$. We also allow for decay of the excited state $\ket{1}$ into the other two states with decay rates $\gamma_{s}$ and $\gamma_{p}$ respectively.  The Hamiltonian for this system is
\begin{align}
\mathscr{H} =  - \hbar (\Omega_p \ket{3}\bra{1}  + \Omega_s(x) \ket{2}\bra{1} ) \re^{- \ri\, \Delta t} + \text{H.~c.}\,.
\label{Eq:Hamiltonian}
\end{align}
The corresponding density matrix equations can be written as
\begin{equation}
\dot{\rho} = \frac{\ri}{\hbar} [\mathscr{H}, \rho] - \sum_{i = 2, 3}\frac{\gamma_{1i}}{2} (\ket{1}\bra{1} \rho - 2 \rho_{11}\ket{i}\ket{i} + \rho \ket{1}\bra{1})\,,
\label{Eq:dotRho}
\end{equation}
with $\gamma_{12}=\gamma_{s}$ and $\gamma_{13} = \gamma_{p}$.
In principle, the steady state solutions of the density matrix equations~\eqref{Eq:dotRho} can be obtained for the given Hamiltonian~\eqref{Eq:Hamiltonian}. Once the steady state is reached, such a system demonstrates coherent population trapping signified by the form of the steady state given by
$\ket{\Psi}=(\Omega_p \ket{2} - \Omega_s(x) \ket{3})/\Omega$, where $\Omega= \sqrt{|\Omega_p|^2 + |\Omega_s(x)|^2}$; this state does not evolve dynamically as $\mathscr{H}\ket{\Psi} = 0$.
Thus, if the atom is initially prepared in the state $\ket{3}$, it will end up in the state $\ket{\Psi}$  at the steady state, as long as the two-photon resonance is maintained. As the population in the state $\ket{\Psi}$ can not escape, it is termed as the trapping state and the phenomena is called coherent population trapping. As shown in our earlier paper~\cite{AgarwalKapale:2006}, taking the drive field to be a standing wave field, $\Omega_s(x)=\Omega_s \sin{k\,x}$, the atomic dynamics becomes position dependent and possesses smaller-than-wavelength ($\lambda = 2 \pi/k$) features at the nodes of the standing wave field as witnessed by monitoring the population of the level $\ket{2}$:
\begin{align}
\rho_{22} = \frac{1}{1 + (\Omega_{s}/\Omega_{p})^2} =\frac{1}{1 + \mathcal{R} \sin ^2{k\,x}}\,.
\label{Eq:Rho22CPT}
\end{align}
This function has the same structure as the transmission function of the Fabry-Perot cavity~\cite{Born:1999}, where the ratio, $\mathcal{R}=|\Omega_{s}|^2/|\Omega_{p}|^2$, of the effective field intensities plays the role of the cavity finesse. The full width at half maximum (FWHM) of the curve is given by $k \Delta x = 2/\sqrt{\mathcal{R}}$. In fig.~\ref{Fig:Levelscheme} (b) we compare the feature-size measurable by the Rayleigh limit ($\lambda/2$) with the one measurable by the 2D version (see Fig.~\ref{Fig:MultiLevel}(a) for the corresponding levelscheme) of the CPT based scheme ($\lambda/\pi \sqrt{\mathcal{R}}$). For a moderate value of $\mathcal{R}=1000$, CPT-based scheme offers resolution improvement by a factor of 50. Further improvement is possible by increasing the value of $\mathcal{R}$.  

\begin{figure}[ht]
\centerline{\includegraphics[width=0.9\columnwidth]{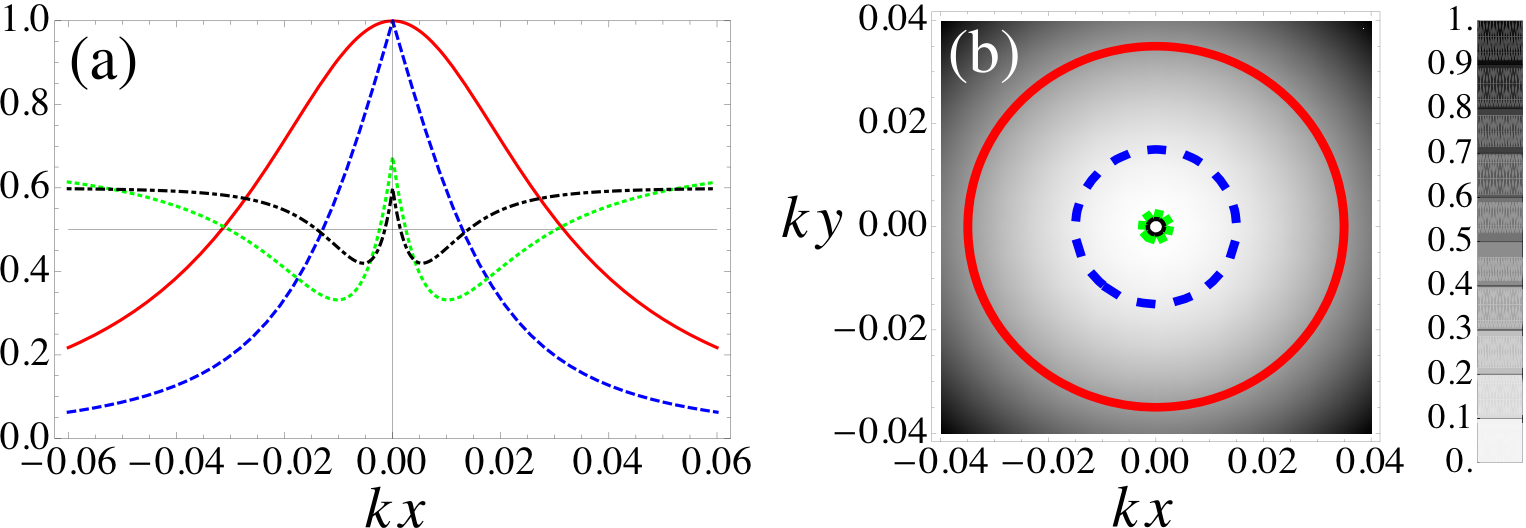}}
\caption{\label{Fig:CPT-DirectMod} (Color online) Fully modulated CPT results with standing wave drive profile. (a) Point spread functions $f_{i}$, Eq.~\eqref{Eq:FMPSF}, vs $kx$ and (b) 2D FWHMs of the point spread functions against the background of the density plot of the 2D standing wave ($\sin^2(kx)+\sin^2(ky)$) near the node. Legend: Red (Solid): unmodulated CPT Result, Blue (Dashed): DC term ($f_{0}$), Green (Dotted): coefficient of $\cos 2\nu t$ ($f_2$) and Black (Dot-Dashed): coefficient of $\cos 4 \nu t$ ($f_4$). The green and black plots are shifted vertically and may be inverted so that the FWHM can be visualized easily.  Here $\mathcal{R}$ is larger by a factor of 2 than the one used in the unmodulated CPT case due to time averaging required for the probe field, that is $\mathcal{R} = 2000$. The resolution improvement from the Rayleigh limit of $\lambda/2$ to the black circle is by a factor of 1000 and the resolution improvement from the red circle (unmodulated CPT result to the black circle (coefficient of the second term in the Fourier expansion) is by a factor of about 23.}
\end{figure}

{\it Full Amplitude Modulation.}---Now we consider full modulation of the probe field with simple standing-wave coupling field and once again monitor the population of level $\ket{2}$, that is $\rho_{22}$ as a measure of the point sperad function. The profiles of the coupling and modulated probe fields can be taken as $
\Omega_{s}(x)$ and $\Omega_{p}(x,t) = \Omega_p(x)\, \cos{\nu t}$.
Thus, the population of the state $\ket{2}$ can be determined to be
\begin{align}
\rho_{22}(t)= \frac{|\Omega_{p}(x)|^2 \cos^2\nu t}{|\Omega_{p}(x)|^2 \cos^2\nu t + |\Omega_{s}(x)|^2}= \sum_{l=1}^{\infty} f_{2l}\cos{(2l\nu t)}\,,
\end{align}
where
\begin{align}
f_{2l}=\sum_{m=\ell}^{\infty} (-1)^{m-1}\(\frac{\Omega_{p}(x)}{\Omega_{s}(x)}\)^{2m} \binom{2m}{m} \frac{2}{2^{2m}} \binom{2m}{m-\ell}\,.
\end{align}
Some of the lower order terms are given by
\begin{align}
f_{0} &= \rho_{22}^{(0)} = 1 - \frac{1}{\sqrt{1+ \[{\Omega_{p}(x)}/{\Omega_{s}(x)}\]^{2}}}\,,\nonumber \\
f_{2} &= \rho^{(1)}_{22} = \frac{2+4 \[{\Omega_{s}(x)}/{\Omega_{p}(x)}\]^{2}}{\sqrt{1+ \[{\Omega_{p}(x)}/{\Omega_{s}(x)}\]^{2}}}  - 4 \[{\Omega_{s}(x)}/{\Omega_{p}(x)}\]^{2}\,,
\nonumber \\
f_{4} &=\rho^{(2)}_{22}=-\frac{2[1+8 \[{\Omega_{s}(x)}/{\Omega_{p}(x)}\]^{2}(1+ \[{\Omega_{s}(x)}/{\Omega_{p}(x)}\]^{2}]}{\sqrt{1+ \[{\Omega_{p}(x)}/{\Omega_{s}(x)}\]^{2}}} \nonumber \\
&\quad\quad + 8 \[{\Omega_{s}(x)}/{\Omega_{p}(x)}\]^{2}(1+2 \[{\Omega_{s}(x)}/{\Omega_{p}(x)}\]^{2})\,.
\label{Eq:FMPSF}
\end{align}
The functions $f_{2l}$, which are functions of the transverse coordinates ($x$ and also of $y$ in the 2D case), are identified as the point spread functions of the proposed scheme for super-resolution. The FWHM of the point sperad function, which is a measure of how a point object is perceived by the method employed for microscopic detection, is a good measure of the resolution.
We now consider the case where the drive field is a standing-wave, $\Omega_{s}(x)=\Omega_{s} \sin(k\,x)$, and the probe field has no position dependence $\Omega_{p}(x) = \Omega_{p}$.  The corresponding point spread functions~\eqref{Eq:FMPSF} are plotted in Fig.~\ref{Fig:CPT-DirectMod} (a), with the general result that the higher-order Fourier components offer better resolution. The scheme can be extended to two dimensions (see Fig.~\ref{Fig:MultiLevel} (a) for the corresponding level scheme) to obtain the plot of the FWHM of the corresponding point spread functions in 2D as in Fig.~\ref{Fig:CPT-DirectMod} (b).
Taking the Fourier transform of the measured fluorescence signal from level $\ket{2}$ one can obtain the term of a certain order. To illustrate, the point spread function corresponding to the $\cos(4 \nu t)$, $f_{4}$, term gives a factor of 1000 improvement over the Rayleigh limit, meaning this scheme can offer resolution of about 0.25 nm.

\begin{widetext}
\begin{figure*}[ht]
\centerline{\includegraphics[width=1.8\columnwidth]{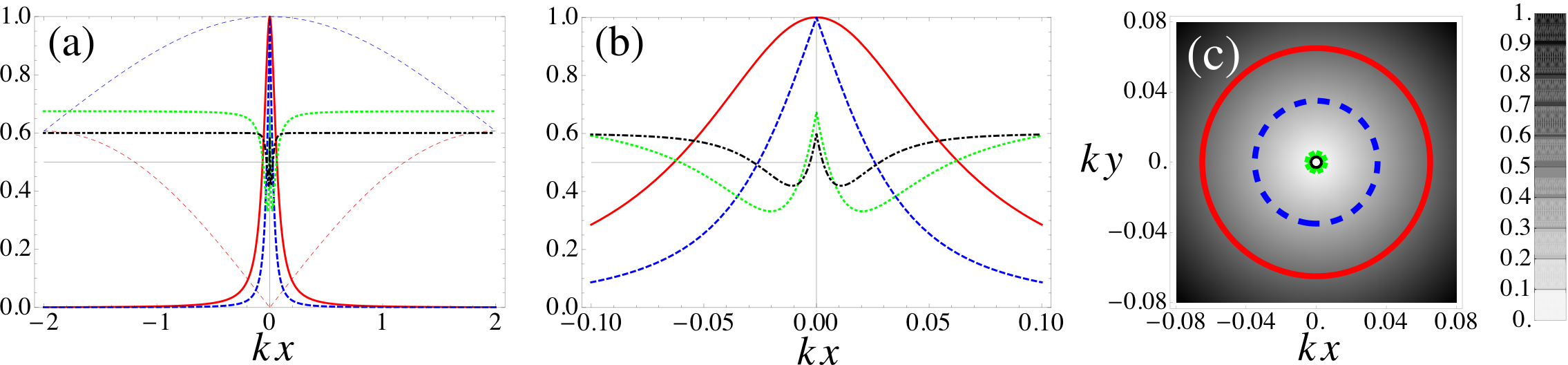}}
\caption{\label{Fig:CPTLGMod} (Color online) Fully modulated CPT Results for transverse Laguerre-Gaussian (LG) drive profile. (a) With transverse drive and probe profile  (b) zoom-in on (a), and  (c) 2D version with the density plot of LG beam in the background. Legend: Red (Solid): unmodulated CPT Result, Blue (Dashed): DC term ($f_{0}$), Green (Dotted): coefficient of $\cos 2\nu t$ ($f_2$) and Black (Dot-Dashed): coefficient of $\cos 4 \nu t$ ($f_4$), and Black (Thin Solid): guide to judge the FWHM. The blue and green plots are shifted vertically for easy comparison of FWHMs of different curves. Here we have chosen $\mathcal{R}=1000$.  It is clear that the dot-dashed black curve corresponding to  $f_{(4)}$ has the sharpest feature and offers the resolution improvement of about 20 compared to the unmodulated CPT result. }
\end{figure*}
\end{widetext}

For certain applications such as microscopy spatially modulated drive and probe fields are more suitable in order to target a smaller area of the sample to begin with. Therefore, in place of the standing wave modulation for the drive field we propose to use a donut (Laguerre-Gaussian) profile for the drive field and a gaussian profile for the probe beam. These profiles used on top of the perturbative modulation can lead to sharpening of the point spread function as discussed below.
The spatial profiles can be written as:
$\Omega_{p}(r) = \re^{- r^2/2 w_{0}^2}$, $\Omega_{s}(r) = \sqrt{\mathcal{R}} \sqrt{{x^2}/{w_{0}^2}}\, \re^{- r^2/2 w_{0}^2}$,
where $w_{0}$ is the beam waist and $r$ is the transverse coordinate. The corresponding point spread functions~\eqref{Eq:FMPSF} are plotted in Fig.~\ref{Fig:CPTLGMod} (a) and (b) along with the beam profiles in one dimension. It is clear that the spatial profiles offered by CPT are much finer than the spatial modulation of the light beams. This scheme can be easily extended to two dimensions in a straightforward manner as the transverse beam profiles are naturally two dimensional.  In the part (c) of the figure Fig.~\ref{Fig:CPTLGMod} we plot the FWHM as the size of the point spread function with the background showing the contour plot of the drive beam profile. The FWHM of the beam profile is too large to fit in the scale shown. When one monitors the second order term, in comparison with the unmodulated CPT results the resolution can be improved by a factor of about 3 and in comparison with the Rayleigh limit by a factor of about 150 for the chosen value of $\mathcal{R}=1000$. Higher values of $\mathcal{R}$, which is the ratio of the intensities of the drive and probe fields offers better resolution.

\begin{figure}[ht]
\centerline{\includegraphics[width=0.9\columnwidth]{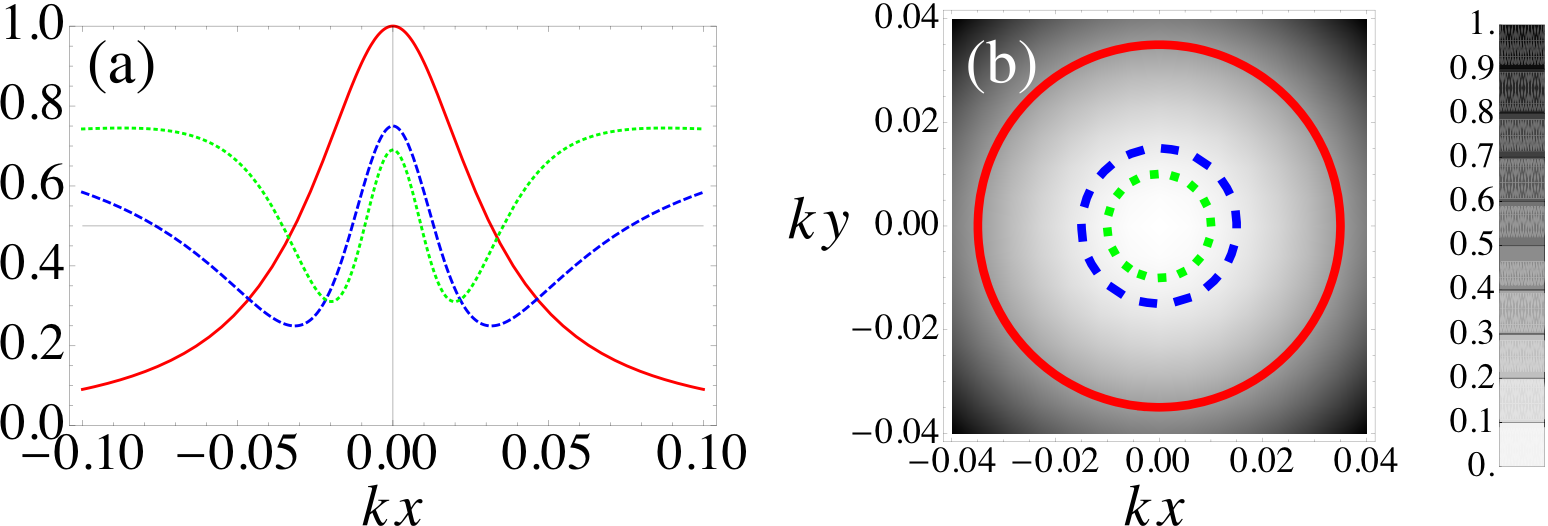}}
\caption{\label{Fig:CPTPMod} (Color online) Perturbatively modulated CPT Results. (a) 1D, and (b) 2D version with the density plot of the 2D standing wave near the node. Legend: Red (solid): Unmodulated CPT Result or $\rho_{22}^{(0)}$, Blue (Dashed): $\rho_{22}^{(1)}$, Green (Dotted): $\rho_{22}^{(1)}$,  and Black (Thin Solid): guide to judge the FWHM. The blue (dashed) and green (dotted) plots are shifted vertically for easy comparison and visualization of FWHM. Here we have chosen $\mathcal{R}=1000$. The resolution improvement for the first order (Green) is by a factor of 3 compared with unmodulated CPT result in Red.}
\end{figure}

{\it Perturbative Amplitude Modulation.}---We conclude the paper by examining a more traditional method of performing modulation spectroscopy. In this case the probe field is amplitude modulated with a perturbative harmonic component $\sin\nu t$ where $\nu$ is the frequency of the harmonic term. Whereas the drive field is spatially modulated along the $x$ direction such that it takes the familiar form of a standing wave field to introduce position dependence in the atomic dynamics. The field Rabi frequencies can be written as:
\begin{align}
\Omega_{s}(x) = \Omega_s\,\sin{k x}\,,\,\,\,
\Omega_{p}(t) = \Omega_p\,(1 + a \sin{\nu t})\,,\,\,\, (a\ll1)\,.
\label{Eq:pert}
\end{align} Using a power series expansion in $a$ for the density matrix elements
$
\rho_{ij} = \rho_{ij}^{(0)} + a \rho_{ij}^{(1)} + a^2 \rho_{ij}^{(2)} + \cdots\,,
$
the steady state solution of the density matrix equations of motion can be  obtained. Once again the population of state $ \ket{2}$ turns out to be position dependent and is given by:
\begin{align}
\rho_{22} &= \[\frac{1}{1 + (\Omega_{s}/\Omega_{p})^2}\] + \[\frac{2 (\Omega_{s}/\Omega_{p})^2}{\left(1+ (\Omega_{s}/\Omega_{p})^2\right)^2}\] (a\,\sin{\nu\,t})  \nonumber \\
&\quad -\[\frac{(\Omega_{s}/\Omega_{p})^2 \left(3- (\Omega_{s}/\Omega_{p})^2\right)}{\left(1+ (\Omega_{s}/\Omega_{p})^2\right)^3}\] (a\sin{\nu\,t})^2 + \cdots\,.
\label{Eq:rho22}
\end{align}
In Fig.~\ref{Fig:CPTPMod} we plot the coefficients enclosed in the square brackets in the above relation, for the field profiles given in Eq.~\eqref{Eq:pert}, versus the dimensionless position coordinate $kx$.  For the current system the point spread function is proportional to the appropriate coefficients determined above. It is clear that the resolution is improved ever so slightly by employing a higher and higher order term in the expansion of population. In an experimental setting, $\rho_{22}$ will be measured as a whole. Then knowing the particular values used for the perturbation parameter $a$ and the harmonic frequency $\nu$ with the help of curve-fitting the required term can be determined by substituting for the lower order terms with their theoretical values.
The perturbative amplitude modulation can also be applied to the case where drive field has the Laguerre-Gaussian beam profile and the probe is Gaussian. Similar results to the case of full modulation are obtained that offer finer resolution than the transverse spatial profile of the beams.

{\it New possibilities.}---
The essentially 1D model presented in Fig.~\ref{Fig:Levelscheme} can be extended to 2D by taking advantage of the multilevel nature of atoms as shown in Fig.~\ref{Fig:MultiLevel}(a). Two standing waves in $x$ and $y$ directions induce sharp features in $\rho_{33}$ allowing sub-nanoscale localization in the $x$-$y$ plane.
Further, considering the lower level states $\ket{2}$ and $\ket{3}$ as hyperfine sublevels of a real atom with $F\geq 2$ one can obtain multiple $\Lambda$ systems in a single atom that are simultaneously excited; such systems have been shown to be useful for generation of skyrmions in the context of BEC and Laguerre-Gaussian beam interaction~\cite{Leslie:2009}. In the case shown in Fig.~\ref{Fig:MultiLevel}(b) the result with unmodulated CPT gets modified to a function that can be approximated by the square of the Fabry-Perot transmission function~\eqref{Eq:Rho22CPT} and hence is sharper by a factor of about 1.5 in the case of unmodulated CPT. Adding the probe field modulation on top of this CPT in the coupled $\Lambda$ system one can, in principle, improve the resolution further. The ideas presented so far can also be extended to three dimensions by employing standing waves in the three directions much on the same lines as techniques for optical trapping of neutral atoms in 3D~\cite{Metcalf:2001}.

{\it Conclusions.}---We offer sub-nanoscale resolution by modulating probe fields in standard CPT. By modulating the CPT  we take advantage of the dispersion accompanying sharp resonances in EIT that is closely related to the phenomena of CPT. The improvement in resolution obtained in comparison with the Rayligh limit is by a factor of about 1000 in the best case scenario discussed here which gives sub-nanometer resolution with the use of optical fields. 

\begin{acknowledgments}
The authors are grateful to the Director of Physical Research Laboratory, Ahmedabad, India for the hospitality where this work was initiated. GSA acknowledges support from NSF grant no. PHYS0653494.
\end{acknowledgments}

\end{document}